\documentclass[12pt]{iopart}
\usepackage{graphicx}
\usepackage{xcolor}
\usepackage{array}
\usepackage{iopams}
\expandafter\let\csname equation*\endcsname\relax
\expandafter\let\csname endequation*\endcsname\relax
\usepackage{amsmath}
\usepackage{siunitx}
\usepackage{bm}
\usepackage{iopams}  
\begin{document}

\title[]{Impact of nuclear effects on the ultrafast dynamics of an organic/inorganic mixed-dimensional interface}

\author{Matheus Jacobs,$^{1}$ Karen Fidanyan,$^{2}$ Mariana Rossi,$^{2}$ and Caterina Cocchi$^{1,3}$} 
\address{$^1$ Humboldt-Universit\"at zu Berlin, Physics Department and IRIS Adlershof, 12489 Berlin, Germany}
\address{$^2$ Max Planck Institute for the Structure and Dynamics of Matter, 22761 Hamburg, Germany}
\address{$^3$ Carl von Ossietzky Universit\"at Oldenburg, Institute of Physics and Center for Nanoscale Dynamics (CeNaD), 26129 Oldenburg, Germany}
\ead{mariana.rossi@mpsd.mpg.de, caterina.cocchi@uni-oldenburg.de}

\begin{abstract}
Electron dynamics at weakly bound interfaces of organic/inorganic materials are easily influenced by large-amplitude nuclear motion. In this work, we investigate the effects of different approximations to the equilibrium nuclear distributions on the ultrafast charge-carrier dynamics of a laser-excited hybrid organic/inorganic interface. By considering a prototypical system consisting of pyrene physisorbed on a MoSe$_2$ monolayer, we analyze linear absorption spectra, electronic density currents, and charge-transfer dynamics induced by a femtosecond pulse in resonance with the frontier-orbital transition in the molecule. The calculations are based on \textit{ab initio} molecular dynamics with classical and quantum thermostats, followed by time-dependent density-functional theory coupled to multi-trajectory Ehrenfest dynamics. We impinge the system with a femtosecond (fs) pulse of a few hundred GW/cm$^2$ intensity and propagate it for 100~fs. We find that the optical spectrum is insensitive to different nuclear distributions in the energy range dominated by the excitations localized on the monolayer. The pyrene resonance, in contrast, shows a small blue shift at finite temperatures, hinting at an electron-phonon-induced vibrational-level renormalization. The electronic current density following the excitation is affected by classical and quantum nuclear sampling through suppression of beating patterns and faster decay times. Interestingly, finite temperature leads to a longer stability of the ultrafast charge transfer after excitation. Overall, the results show that the ultrafast charge-carrier dynamics are dominated by electronic rather than by nuclear effects at the field strengths and time scales considered in this work. 
\end{abstract}

%\maketitle

\newpage

\section{Introduction}
Low-dimensional hybrid interfaces formed by carbon-conjugated molecules physisorbed on semiconductor transition-metal dichalcogenide (TMDC) mono- and few-layers represent a new material class with intriguing properties in the field of optoelectronics~\cite{choi+16nano,guo+22nr}.
The characteristics of the resulting system largely depend on the choice of the constituents and in particular on the nature of the molecules~\cite{cai+16cm,amst+19nano,koch21apl,park+21as}.
Organic donor and acceptor moieties can give rise to strong electronic interactions with the underlying semiconductor TMDCs and, thus, to charge-transfer phenomena already in the electronic ground state~\cite{Park2021,zhen+16nano,wang+18afm}, usually related to the alignment of the frontier states and their hybridization~\cite{wang+19aelm,habi+20ats,chen+21jpcl,chris+prb2023}.
On the other hand, purely carbon-based molecules such as pyrene, perylene, and larger polycyclic aromatic hydrocarbons typically bind weakly to the substrate~\cite{krum-cocc21es,mela+22pccp,krum-cocc23pssa,tand+23pssa}.
As a consequence, a high degree of hybridization of their frontier orbitals with the TMDC electronic bands can only take place under certain conditions~\cite{krum-cocc21es}. 

To properly assess the potential of such systems for optoelectronic applications, it is necessary to explore their properties beyond the electronic ground state, for example, by investigating the nature of their excitons~\cite{oliv+22prm,thom+23ns} and the response of the charge density to coherent perturbations~\cite{jacobs+ACS2022}.
\textit{Ab initio} methods such as real-time time-dependent density functional theory (RT-TDDFT) are suitable for this task as they enable the simulation of the material and the evolution of its electron density without the need for empirical models~\cite{bert+00prb,taki+07jcp}.
Established developments of this approach allow including explicitly external, time-dependent fields in the calculations~\cite{lopa-govi11jctc,degi+13chpch}, thus effectively mimicking the scenario of pump-probe experiments~\cite{krum+20jcp,uemo+21prb,yang+22cpl}.
In a recent study based on RT-TDDFT~\cite{jacobs+ACS2022}, some of the authors disclosed the mechanisms ruling the response of the interface between pyrene and monolayer MoSe$_2$ to a femtosecond pulse in resonance with the electronic transition between the frontier states (HOMO and LUMO) of the molecule: When the intensity of the incoming laser field is weak enough so as to remain in the linear-response regime,
%such that the response of the system is linear,
pyrene acts as a donor and the TMDC as an acceptor. However, as the intensity of the incoming field increases and the response of the system enters the nonlinear regime, Pauli-blocking comes into play and reverts the direction of charge transfer, making the monolayer electron-donating and the molecule electron-withdrawing. 
In that work~\cite{jacobs+ACS2022}, only very short time scales (30~fs) were considered, and the nuclear motion was ignored.
%, as it does not come into play in the ultrashort timescale of 30~fs considered therein.
%On the other hand, on a temporal window of 100~fs or longer, these effects cannot be neglected. 
While the coupled electron-nuclear motion was shown to be crucial in the ultrafast dynamics of organic blends~\cite{falke+Sci14,jaco+23jpca} as well as in prototypical hybrid systems characterized by high levels of doping~\cite{jaco+20apx}, its influence on weakly bound organic-inorganic interfaces is still largely unexplored.

When dealing with the nuclear motion of materials excited by ultrafast and coherent radiation, the Ehrenfest molecular dynamics scheme is a popular choice that enables the description of electron-vibrational couplings including non-adiabatic effects~\cite{rozz+17jpcm,birc+17sd}.
Its interface with RT-TDDFT has been well established for two decades~\cite{marques2003octopus}.
However, most applications still rely on the single-trajectory scheme, in which the nuclear distribution is represented by a single, arbitrary configuration.
Recent work~\cite{lively21jpcl,krum+22prb,live+23arxiv} demonstrated the need to go beyond the single-trajectory approach in order to achieve a comprehensive and realistic description of the nuclear motion and its (de)coherence effects when the system is subject to an external perturbation.
Multi-trajectory Ehrenfest, considering the evolution of the nuclear degrees of freedom according to a certain ensemble distribution, has been proven to yield a superior description of nuclear effects in combination with different electronic-structure methods~\cite{lively21jpcl,live+23arxiv,hoff+19jcp,chow+21jcp,choi-vani21jcp,rune-mano23arxiv}.
In particular, averaging multiple trajectories offers a more realistic picture of the decoherence effects in molecules perturbed by resonant pulses~\cite{krum+22prb}.
The nuclear ensembles can be constructed by means of classical or quantum distributions including harmonic and anharmonic effects~\cite{koss-barb18jctc,ross+18jcp}. %within a harmonic approximation or going beyond it~\cite{something}.
Understanding the effects of different approximations for the nuclear motion on the ultrafast electron dynamics of a laser-excited system is particularly important to achieve a reliable description of the physics in play. 
%Low-dimensional hybrid interfaces are ideal platforms for such a study, as they enable the assessment of nuclear effects and the distribution of their degrees of freedom in the dynamical charge transfer between the organic and the inorganic subsystems. 

In this work, we focus on the interface formed by pyrene physisorbed on a MoSe$_2$ monolayer and investigate the effects of different approximations for the equilibrium nuclear distribution and of nuclear motion on its charge carrier dynamics. 
Specifically, we consider the system coherently excited by an ultrafast laser pulse in resonance with the lowest-energy excitation of the molecule.
We apply RT-TDDFT in conjunction with a multi-trajectory Ehrenfest scheme considering initial equilibrium nuclear distributions created by classical Boltzmann sampling and by quantum thermostat sampling.
We analyze the effects of these choices first on the linear absorption spectrum and, subsequently, on the current density and the dynamical charge distribution among the components of the interface, following the excitation with a resonant pulse.

%%%%%%%%%%%%%%%%%%%%%%%%%%%%%%%%%%
\section{Methodology}

\subsection{Theoretical Background}

For a system with coupled electronic and nuclear degrees of freedom, one can derive Ehrenfest dynamics based on a density-matrix formalism. In this formulation, previously described elsewhere~\cite{grun+spring2009}, one obtains coupled equations of motion for the electronic and the nuclear densities. It becomes clear that in a trajectory-based approach, only an ensemble of trajectories for the nuclei can satisfy the solution of the coupled equations of motion providing a rigorous basis for multi-trajectory Ehrenfest methods. 

The expectation value of a time-dependent observable in this formulation can be written as~\cite{geras+tmp1982}
 \begin{equation}
 \langle \hat{A}(t) \rangle = \frac{1}{Z} \Tr' \left[ \hat{\rho}_e \int d\bm{P} d\bm{R}  \rho^{\beta}_{n}(\bm{R}, \bm{P}) \hat{A}_W(\bm{R}(t), \bm{P}(t)) \right],
 \end{equation}
 where the primed trace denotes a trace only over the electronic degrees of freedom, $\hat{\rho}_e$ is the electronic reduced density matrix operator, the subscript $W$ indicates a partial Wigner transform over the degrees of freedom of the nuclei, $\bm{R}$ and $\bm{P}$ are Wigner-transform variables corresponding to the nuclear position and momenta, and $\rho_n^\beta$ is the nuclear canonical density matrix at an inverse temperature $\beta$, defined as $\rho_n^{\beta} = \text{Tr}'[\hat{\rho}_W^{\beta}]$. This is a semiclassical mean-field formulation, where we used the Wigner transform of the full density matrix $\hat{\rho}^{\beta}_W = (e^{-\beta \hat{H}})_W \approx \hat{\rho}_e \rho^{\beta}_{n}$. The last assumption means that correlations between the electronic and nuclear densities are disregarded. As long as it is possible to sample $N$ initial conditions from $\rho^{\beta}_{n}(\bm{R}, \bm{P})$, one can calculate the phase-space average by
  \begin{equation}
 \langle \hat{A}(t) \rangle = \frac{1}{N} \sum_{i=1}^N \Tr' \left[ \hat{\rho}_e \hat{A}_W(\bm{R}_i(t), \bm{P}_i(t)) \right].
 \end{equation}
The protocol employed in this paper is to first sample $N$ initial conditions $\bm{R}_0$ and $\bm{P}_0$ (for the nuclear degrees of freedom) from an approximation of $\rho^{\beta}_{n}$ and then evolve the observables in time according to Ehrenfest dynamics.

% Note that when dealing with the canonical ensemble, it should be sampled for the initial conditions and then one can show that for a large amount of particles and trajectories, the probability density in the ``microcanonical'' ensemble (which is generated by the time evolution of the trajectories) approaches that of the canonical ensemble.

% {\color{purple} Complete here with better analysis and better text}

% %Much in the same spirit of classical thermodynamics, invoking the ergodic hypothesis tells us that this phase-space average can also be calculated by
% %\begin{equation}
% %    \langle \hat{A} \rangle = \lim_{T \to \infty} \frac{1}{T} \int_0^T dt \, \hat{A}_W(Q(t), P(t))
% %\end{equation}
% %if (i) one knows how to evolve $Q$ and $P$ in time, in a form consistent with the transformed Hamiltonian that enters $\hat{\rho}^{\beta}_W(Q, %P)$, (ii) the initial conditions $Q_0$ and $P_0$ are drawn from $\hat{\rho}^{\beta}_W(Q, P)$ and (iii) the trajectories followed by $Q(t)$ and %$P(t)$ conserve $\hat{\rho}^{\beta}_W(Q, P)$ (this is not true in mean-field Ehrenfest!). Only the last point we don't have (ask). 

\subsubsection{Classical-nuclei sampling}

 When doing the sampling with classical nuclei, we are effectively making the following approximation:
 \begin{equation}
     \rho^{\beta}_{n}(\bm{R}, \bm{P}) = e^{-\beta H(\bm{R}, \bm{P})},
 \end{equation}
\textit{i.e.}, we are approximating the Wigner-transformed nuclear part of the canonical density distribution by a completely classical Boltzmann distribution. While such approximation goes beyond the frequently employed harmonic approximation~\cite{lively21jpcl, krum+22prb} in terms of including correlations and anharmonicity between the different nuclear degrees of freedom, it completely neglects nuclear quantum effects.

 \subsubsection{An approximation to quantum-nuclei sampling}

 As an attempt to include nuclear quantum effects while maintaining the complete treatment of anharmonicity for the nuclear degrees of freedom, we have employed a quantum thermostat~\cite{ceri+prl2009,ceri+10jctc}.  The quantum thermostat
 generates
 a Gaussian stationary distribution for the position and momenta $\rho^{QT}(\bm{R}, \bm{P})$ which is
 consistent with the harmonic quantum mechanical expectation values of $\langle R^2 \rangle$ and $\langle P^2 \rangle$,
 \begin{equation}
 \langle R^2 \rangle=\frac{\hbar}{2m\omega} \coth \frac{\beta\hbar\omega}{2}\quad
 \langle P^2 \rangle =\frac{\hbar m\omega}{2} \coth \frac{\beta\hbar\omega}{2},
  \label{eq:exp-q2-p2}
 \end{equation}
 with no cross-correlations between position and momentum (\textit{i.e.}, $\langle PQ \rangle=0$). 
 In this way, we are thus approximating
 \begin{equation}
     \rho^{\beta}_{n}(\bm{R}, \bm{P}) = \rho^{QT}(\bm{R}, \bm{P}).
 \end{equation}
 This equation approximates nuclear quantum effects by enforcing an effective distribution of position and momenta that mimics the quantum distribution. It is rigorous if the system is completely harmonic (comparable to harmonic Wigner sampling). Through the molecular dynamics sampling of the system potential with this thermostat, anharmonicities are introduced. The protocol has been shown to work well in these cases~\cite{druz+jpcs2018}. We employ this approximation here as an uncontrolled approximation to the nuclear quantum distribution, with the sole purpose of probing the potential effect of quantum statistics on the initial distribution of nuclear degrees of freedom. 

\subsubsection{Calculation of optical responses}

To calculate the optical response of the pyrene@MoSe$_2$ interface to an ultrafast laser field, we performed first-principles simulations in the framework of RT-TDDFT coupled with Ehrenfest molecular dynamics. 
In this formalism, the time-dependent Kohn-Sham (KS) equations~\cite{rung-gros84prl}
\begin{equation}
i\frac{\partial}{\partial t}\phi_{n\mathbf{k}}(\textbf{r},t) =\left( -\frac{\nabla_{\mathbf{r}}^2}{2}+v_{\text{KS}}[n](\textbf{r},t)\right)\phi_{n\mathbf{k}}(\textbf{r},t),
\label{eq:TD-KS}
\end{equation}
are solved within the adiabatic approximation for the exchange-correlation term ($v_{xc}$) included in the effective KS potential $v_{\text{KS}}[n]$. In these expressions, $n$ is the electronic density, and $\bm{r}$ represents electronic positions. The velocity-gauge formulation is assumed, including an external electric field of the form $\textbf{E}(t) = -\dfrac{1}{c}\dfrac{d\textbf{A}(t)}{dt}$, where $\textbf{A}$ is the vector potential and $c$ the speed of light in vacuo.
The output quantity relevant to the analysis presented in this work is the time-dependent macroscopic current density computed as
\begin{equation}\label{eq.current}
    \textbf{J}(t) = \int_{\Omega} \text d^{3}r~\frac{1}{2}\sum_{n\textbf{k}}  \left[\phi_{n\textbf{k}}^{*}(\textbf{r},t)\left(-i\nabla + \frac{\textbf{A}(t)}{c} \right)\phi_{n\textbf{k}}(\textbf{r},t) + c.c. \right],
\end{equation}
where $\Omega$ is the volume of unit-cell and $\phi_{n\textbf{k}}$ are the solutions of Eq.~\eqref{eq:TD-KS}.

A partial charge analysis in the framework of the Bader scheme~\cite{bader90} is performed on top of the calculated time-dependent electron density in order to evaluate the amount of charge transfer at the interface. 

%%%%%%%%%%%%%%%%%%%%%%%%%%%
\subsection{Computational Details}
To simulate the various nuclear ensembles we performed \textit{ab initio} molecular dynamics (AIMD) simulations of the pyrene molecule on a MoSe$_2$ monolayer with the FHI-aims code~\cite{blum+09cpc} interfaced to the i-PI code~\cite{kapi+19cpc}. The simulation cell consists of 4$\times$4 the unit cell of isolated MoSe$_2$ and includes one molecule which is separated by its in-plane replicas by about 5~\AA{}.
In the out-of-plane direction, an amount of vacuum equal to 60 \AA~ is included to prevent spurious interactions among the replicas. In addition, a dipole correction is applied. We employed the local-density approximation (LDA) with a $3\times3\times1$ \textbf{k}-point grid and the standard \textit{light} settings of FHI-aims. 
In previous work~\cite{jacobs+ACS2022}, we checked that for the pyrene@MoSe$_2$ interface under study, LDA is able to deliver a description of the electronic structure and in particular of the level alignment between the molecule and the TMDC in qualitative agreement with the one provided by a range-separate hybrid functional~\cite{krum-cocc21es}.
The AIMD simulations were run at the constant temperature $T=100$~K, as the molecule easily desorbs at higher temperatures. We applied the stochastic velocity-rescaling thermostat~\cite{bussi+jcp2007} with a relaxation time $\tau=40$~fs and the generalized-Langevin equation quantum-thermostat with 6 fictitious degrees of freedom and $\hbar \omega / kT = 50$ ~\cite{ceri+prl2009}. A time-step of 0.5~fs was used to integrate the equations of motion. After 5~ps of thermalization, we picked 20 snapshots separated by 500 fs from the classical-nuclei and the quantum-thermostat simulations. The position and momenta corresponding to each one of these snapshots were given as input to the Ehrenfest RT-TDDFT simulations. 

All RT-TDDFT calculations were performed with the OCTOPUS code~\cite{tan-dej2020jcp}. To simulate the effectively two-dimensional pyrene@MoSe$_2$ interface, a \textbf{k}-grid with 3$\times$3$\times$1 points was adopted, and a Coulomb cutoff of 20~\AA{} applied in the $z$-direction. The wave functions were represented on a cubic grid with spacing 0.17~\AA. We used the adiabatic LDA (ALDA) kernel and employed the Hartwigsen-Goedecker-Hutter-LDA pseudopotentials~\cite{hart+98prb} to treat core electrons. 
Each selected nuclear snapshot was propagated for 100~fs with Ehrenfest dynamics coupled with RT-TDDFT with a time-step of 0.0072~fs using the approximated enforced time-reversal symmetry propagator~\cite{cast+04jcp}. The time-dependent electric field is modeled as an $x$-polarized pulse with a Gaussian envelope centered at $t_\mu=12$~fs, with standard deviation $t_\sigma=2$~fs, and carrier frequency $\omega_0 = 3.1$~eV in resonance with the transition between the highest occupied molecular orbital (HOMO) and the lowest unoccupied molecular orbital (LUMO) of pyrene. The intensity of the incoming electric field is set to $2\times10^{11}$~W/cm$^{2}$. In a previous study, this value was shown to maintain the system in the linear regime~\cite{jacobs+ACS2022} of the electronic response. In the Bader charge analysis~\cite{bader90,bader_atoms94,henkelman+06CMS}, the time-dependent density was printed at intervals of 0.72~fs and analyzed by summing the individual charges of each atom belonging to each subsystem (MoSe$_2$ and pyrene).

%%%%%%%%%%%%%%%%%%%%%%%

\section{Results and Discussions} 

\begin{figure}
    \centering
\includegraphics[width=0.9\textwidth]{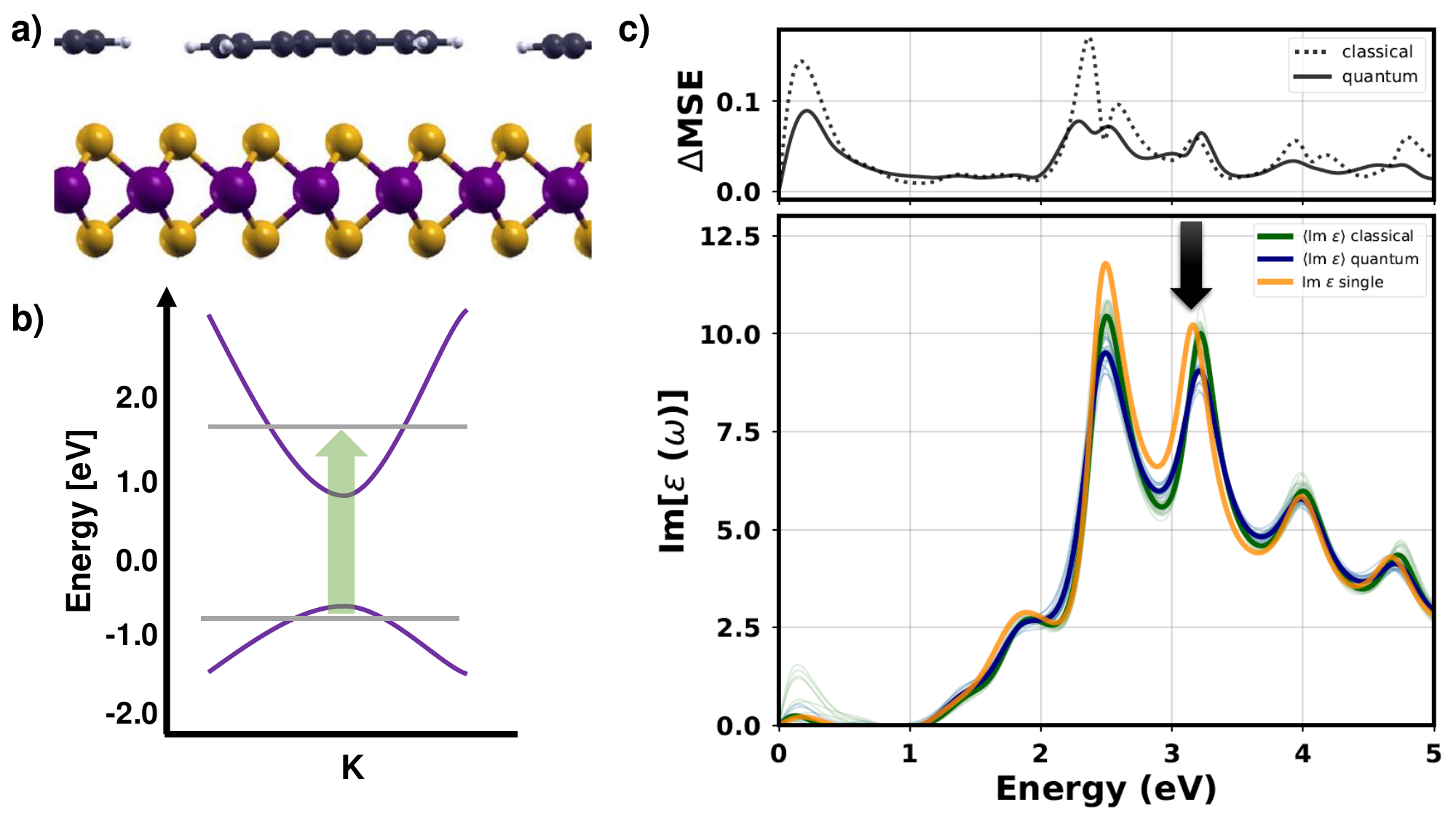}
\caption{a) Ball-and-stick representation (side view) of the pyrene@MoSe$_2$ interface considered in this work: C atoms are shown in black, H in white, Mo in purple, and Se in yellow. b) Schematic level alignment of the pyrene@MoSe$_2$ interface based on the DFT results reported in Ref.~\cite{jacobs+ACS2022} with the bands of MoSe$_2$ shown in purple (valence band maximum and conduction band minimum at the high-symmetry point K) and the frontier levels of the molecule in gray; the green arrow marks the HOMO$\rightarrow$LUMO transition in pyrene targeted by the applied pulse. c) Linear absorption spectra of pyrene@MoSe$_{2}$ considering the optimized structure (orange) and resulting from averages over the classical (green) and quantum (blue) nuclear distributions. The thin curves in the background show the result obtained for each trajectory of all distributions before the averaging. The black arrow marks the excitation targeted by the resonant pulse. The upper panel shows the mean square error for the classical and quantum distribution averaged over 20 snapshots.}
    \label{fig:spectra}
\end{figure}

The system considered in this work is the low-dimensional hybrid interface formed by pyrene molecules physisorbed on a MoSe$_2$ monolayer (see Fig.~\ref{fig:spectra}a).
The electronic structure of this system, extensively analyzed in Refs.~\cite{krum-cocc21es,mela+22pccp,jacobs+ACS2022}, is schematically depicted in Fig.~\ref{fig:spectra}b. 
The two constituents exhibit a type-I level alignment with the frontier states of the TMDC comprised between the HOMO and the LUMO of the molecule. 
Note, however, the peculiar proximity between the HOMO of pyrene and the valence band maximum of MoSe$_2$, which is expected to favor hole transfer in the dynamics. 
The imaginary part of the dielectric function of the optimized structure is characterized by an onset starting slightly above 1~eV, with a first maximum at about 2~eV followed by two sharper resonances at 2.5 and 3.1~eV (see Fig.~\ref{fig:spectra}c, orange curve). 
We mention in passing that the maximum that is visible at about 0.1~eV is a known numerical artifact of RT-TDDFT due to the finite duration of the propagation~\cite{yama-yaba19prb}. 
The first two peaks at about 2.0 and 2.5~eV are excitations stemming from the electronic structure of the TMDC~\cite{jacobs+ACS2022} and in particular from the interband transitions at the K (Fig.~\ref{fig:spectra}b) and $\Gamma$ valleys~\cite{krum-cocc21es}.
Due to the absence of explicit electron-hole interactions in the adopted formalism, their excitonic nature~\cite{rama+12prb} cannot be captured. 
The absorption maximum at 3.1~eV (see Fig.~\ref{fig:spectra}c) corresponds to the HOMO$\rightarrow$LUMO transition in pyrene when physisorbed on MoSe$_2$ (green arrow in Fig.~\ref{fig:spectra}b and black arrow in Fig.~\ref{fig:spectra}c) according to the results of previous work~\cite{jacobs+ACS2022}. 
In the subsequent ultrafast dynamics, we target this excitation by setting the pulse in resonance with it. 

Before proceeding with this analysis, we inspect the linear absorption spectrum of the interface resulting from an average over classical and (approximate) quantum nuclear initial distributions (CD and QD, respectively) obtained with the quantum thermostat. 
The linear absorption spectra shown in Fig.~\ref{fig:spectra}c are obtained as an average over 20 simulations starting from different initial configurations with distributions thermalized at $T= 100$ K.
As shown in Fig.~\ref{fig:spectra}c, the results obtained by averaging the corresponding contributions resemble closely the outcome of the calculation starting from the optimized structure. In the upper panel, we display the mean square error ($\Delta$MSE) obtained from averaging the CD and QD, which is a measure of the statistical error on the spectral average.
The main spectral features described above are largely reproduced. 
Differences appear in the relative height of the peaks, especially at 2.5~eV, where the CD and QD lead to lower oscillator strengths, with QD yielding the lowest intensity.
Moreover, a blueshift of the order of 60-80~meV manifests itself in the resonance at 3.1~eV, consistent with a renormalization of the HOMO-LUMO gap of pyrene due to adiabatic electron-phonon coupling.
At higher energies, the three spectra shown in Fig.~\ref{fig:spectra}c essentially overlap. 
We explain this behavior by recalling that in the low- and high-energy part of the spectrum ($E<2$~eV and $E>3.5$~eV, respectively), the optical transitions involve mainly, when not exclusively, states of MoSe$_2$~\cite{jacobs+ACS2022}, which, due to the rigidity of this material, are almost insensitive to structural changes at this temperature. 
On the other hand, the peak at 3.1~eV stemming from the HOMO$\rightarrow$LUMO transition of pyrene is more sensitive to conformational changes.

\begin{figure}
    \centering
    \includegraphics[width=0.7\textwidth]{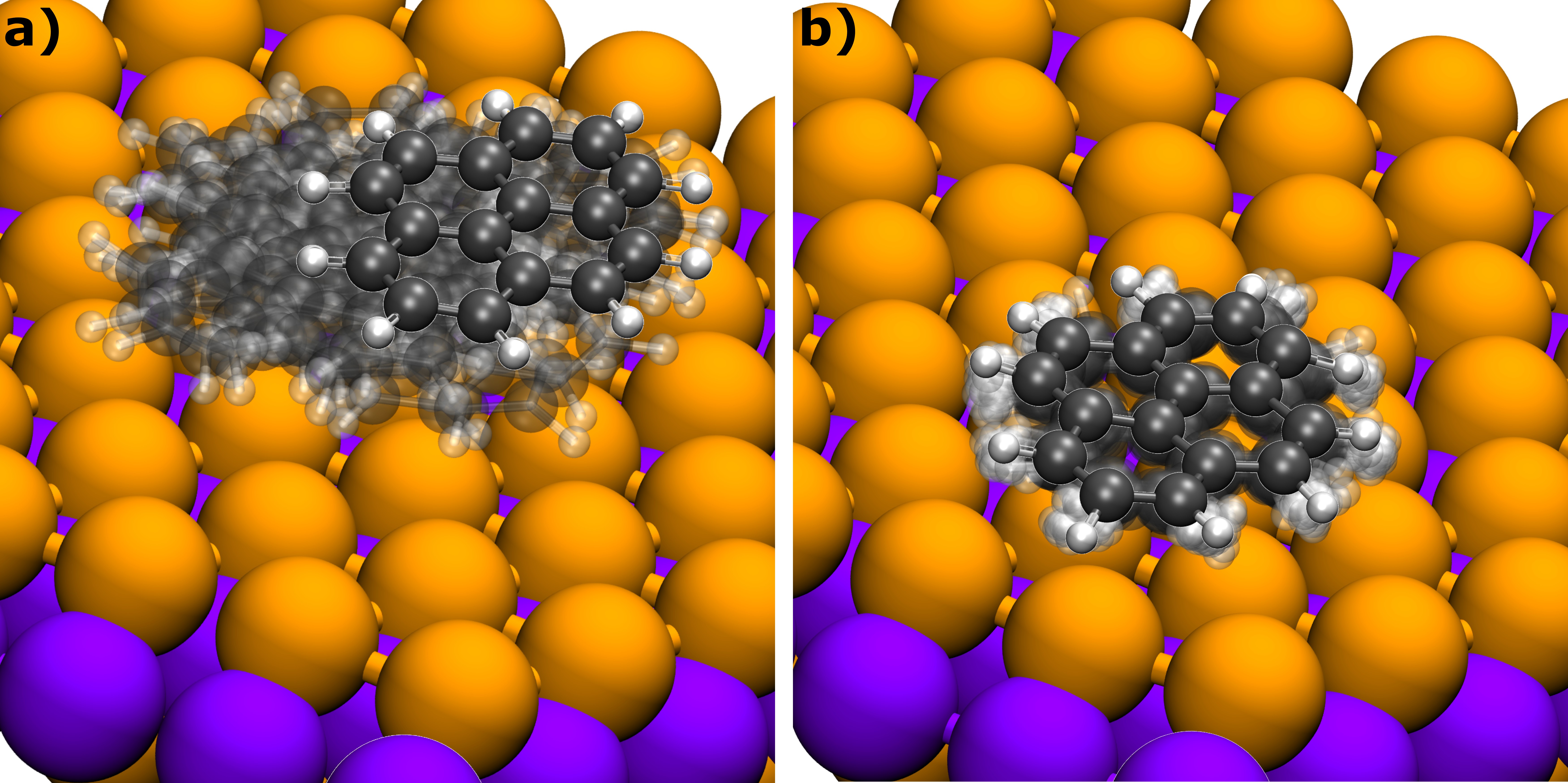}
    \caption{Initial configurations of pyrene@MoSe$_2$ obtained from AIMD a) with classical nuclei at a temperature $T=100$ K and b) with a quantum thermostat at the same temperature. These structures were used to initialize the calculations of the optical spectra and the RT-TDDFT+Ehrenfest simulations. The minimum energy structure is highlighted. \label{fig:snaps}}
\end{figure}

To better understand the differences in the spectra computed from CD and QD, it is instructive to inspect the corresponding starting structures.
Analyzing the initial ensemble of geometries from the MD simulations, shown in Fig.~\ref{fig:snaps}, the most pronounced displacements at $T = 100$~K correspond to translational and hindered rotation modes between the molecule and the TMDC substrate in the classical simulation (Fig.~\ref{fig:snaps}a). In contrast,  the most pronounced displacements in the quantum-thermostat simulation stem from internal molecular distortions and only slight hindered rotations (Fig.~\ref{fig:snaps}b). The lack of (hindered) diffusion of the molecule in the simulation with the quantum thermostat can be caused by the impact that the thermostat has on the nuclear dynamics, which can slow down diffusive degrees of freedom~\cite{ceri+10jctc}. Nevertheless, we note that the average distance between the molecular center of mass and the top layer of Se atoms is 3.24~\AA{} in the quantum case and 3.34~\AA{} in the classical case. This raises the possibility that the ``quantum molecule'' may be more strongly bound to the substrate than the ``classical molecule'', but further investigations on this topic are beyond the scope of this paper. As we will see next, these differences are not crucial for the analysis of electronic excitations. 

The marginal differences induced in the optical spectra by the CD and QD initial sampling, despite the notable structural variations discussed above, are worth further analysis. Structural variations could alter the pyrene-MoS$_2$ hybridization~\cite{krum-cocc21es}, but we do not seem to observe this effect, which can be explained by the following considerations. The most intense resonances that are visible in Fig.~\ref{fig:spectra}c are due to excitations localized within the individual constituents and thus, are rather insensitive to variations of their mutual arrangement.
Hybrid excitons were identified by Oliva and coworkers in the low-energy part of the spectrum of the related heterostructure pyrene@MoS$_2$~\cite{oliv+22prm}.
These features, however, are characterized by very low oscillator strength due to the negligible overlap between the electron and hole localized on the TMDC and on the molecule, respectively.
In the theoretical framework adopted in this work, an equally detailed analysis of the excitations~\cite{oliv+22prm} is not possible and, hence, changes to hybrid excitons induced by thermally-induced molecular reorientations are not detectable.
A dedicated study using the Bethe-Salpeter-formalism~\cite{onida2002rmp,vorwerk2019bethe} will be necessary to clarify this point.

\begin{figure}[h!]
    \centering
    \includegraphics[width=0.9\textwidth]{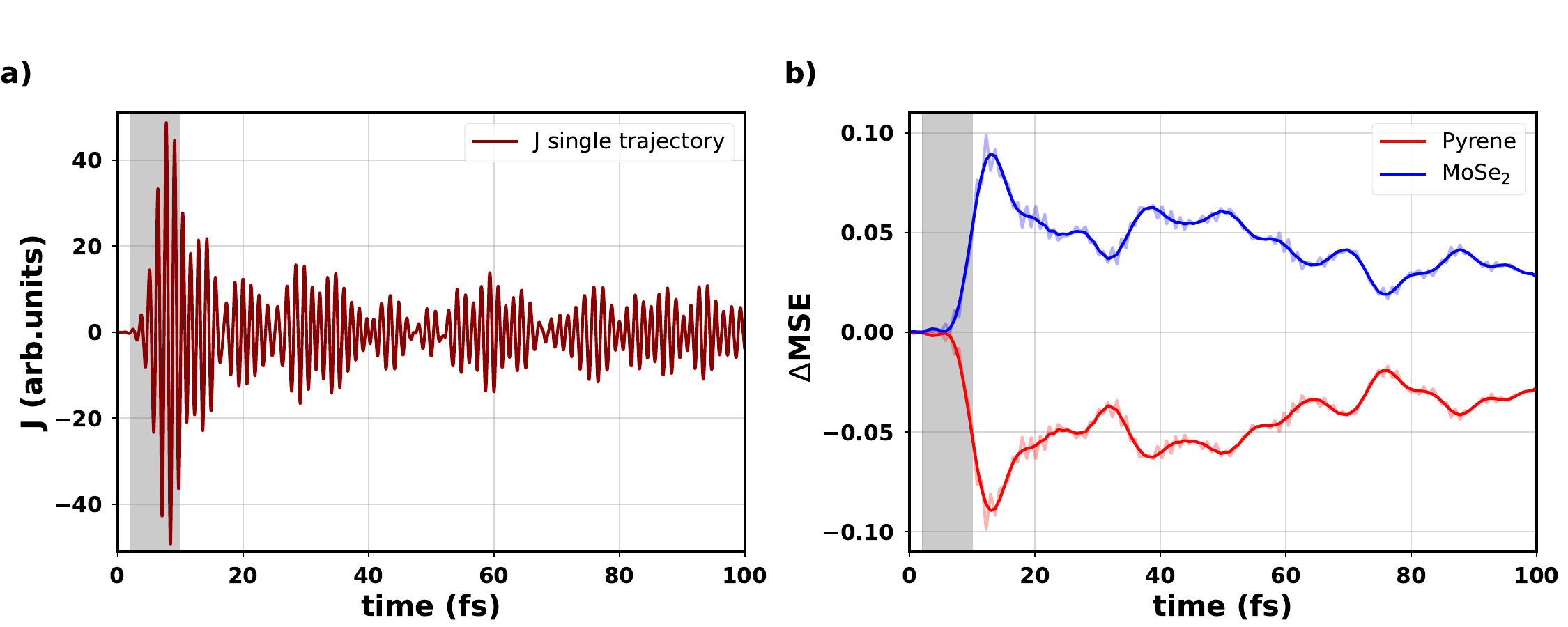}
    \caption{ a) Electronic current density for a single trajectory starting from the minimum-energy structure b) Time-dependent variation with respect to the ground-state value of the Bader partial charges on pyrene and MoSe$_2$ for the same single trajectory as in a).  Solid lines represent the smoothed data using the Savitzky-Golay filter, as for fading curves in the back represent the raw data. Positive (negative) values indicate electron accumulation (depletion). In both panels, the gray area indicates the time window in which the pulse is active. }
    \label{fig:0K}
\end{figure}

Before diving into the discussion of the dynamics of the hybrid interface considering different nuclear ensembles, we inspect the results obtained from a typical single-trajectory RT-TDDFT Ehrenfest dynamics starting from the optimized geometry (see Fig.~\ref{fig:0K}).
We start with the analysis of the current density (Fig.~\ref{fig:0K}a), representing the response of the system impinged by the resonant pulse, where we notice significant oscillations from positive to negative values with the maximal amplitude found immediately after turning off the pulse.
Subsequently, the nuclear motion leads to a relatively slow damping of the signal, where oscillations in $J$ continue to be noticeable. 
The continuous beating pattern in $J$ indicates the presence of active excited states during the whole simulation. 
Moving now to the time-dependent density distribution across the two subsystems (Fig.~\ref{fig:0K}b), we find charge accumulation (depletion) on MoSe$_2$ (pyrene) when the heterostructure is excited with the resonant pulse of frequency $\omega_0=3.1$~eV, in agreement with the results obtained in the absence of nuclear motion under the action of the same pulse~\cite{jacobs+ACS2022}.
In this single-trajectory description, the main effect induced by the nuclear motion, dominated in this time window by the modes for the pyrene carbon network~\cite{herp+21jpca}, is to dampen the amount of charge transfer over time.

\begin{figure}[h!]
    \centering
    \includegraphics[width=\textwidth]{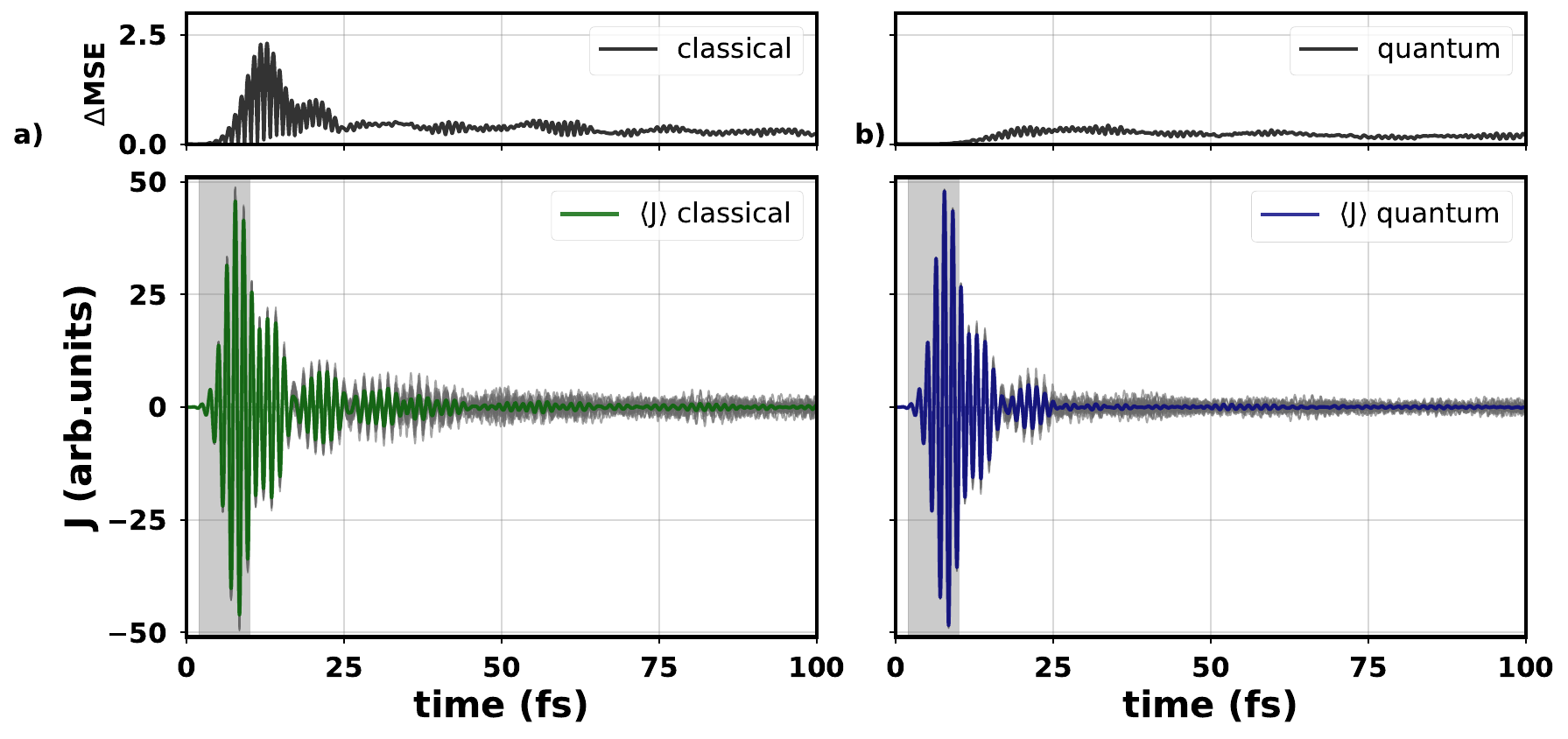}   
     \caption{Electronic current density averaged over 20 snapshots from a) CD and b) QD. The gray signals in the background represent the raw results before the averaging. The gray area indicates the time window in which the pulse is active. The upper panels display the mean-square errors associated with the a) CD and b) QD. }
   \label{fig:current}
\end{figure}

After this digression, we are equipped for the discussion of the nuclear effects simulated in the multi-trajectory approach with CD and QD.
We start by inspecting the electronic current density (see Fig.~\ref{fig:current}).
In comparison to Fig.~\ref{fig:0K}, we notice in this case a much more pronounced damping of $J$ over time after the pulse is switched off. 
Similar behavior was found also in the transition dipole moment of small molecules~\cite{krum+22prb} and can be ascribed to the dissipation induced in the system by the thermalized nuclear degrees of freedom, which otherwise would preserve an unrealistic over-coherence.
The beating pattern already discussed in Fig.~\ref{fig:0K}a persists to a certain degree in the thermalized ensembles, too. 
Despite their general similarity, the results obtained with the nuclear CD and QD exhibit some non-negligible differences. In the classical ensemble (Fig.~\ref{fig:current}a), there are residual oscillations left by the incident field even after it is turned off, with the current density oscillating at the same frequency as the laser pulse. These residual oscillations indicate the presence of laser-induced excited states, presenting a beating pattern that decays to almost zero over the first 50~fs. 
The current density obtained from the QD (Fig.~\ref{fig:current}b) is very similar to the one obtained from the CD during the first 15~fs. Afterward, when the pulse is turned off, the decay in $J$ is much more pronounced: only a single beat can be seen until 25~fs, then, the signal is almost completely damped. 
These results suggest that the decay mechanism and lifetime of the excited states may change when considering temperature effects with CD or QD, but a more extensive analysis in this direction is still needed to clarify this point in full.

\begin{figure}[h!]
    \centering
    \includegraphics[width=.75\textwidth]{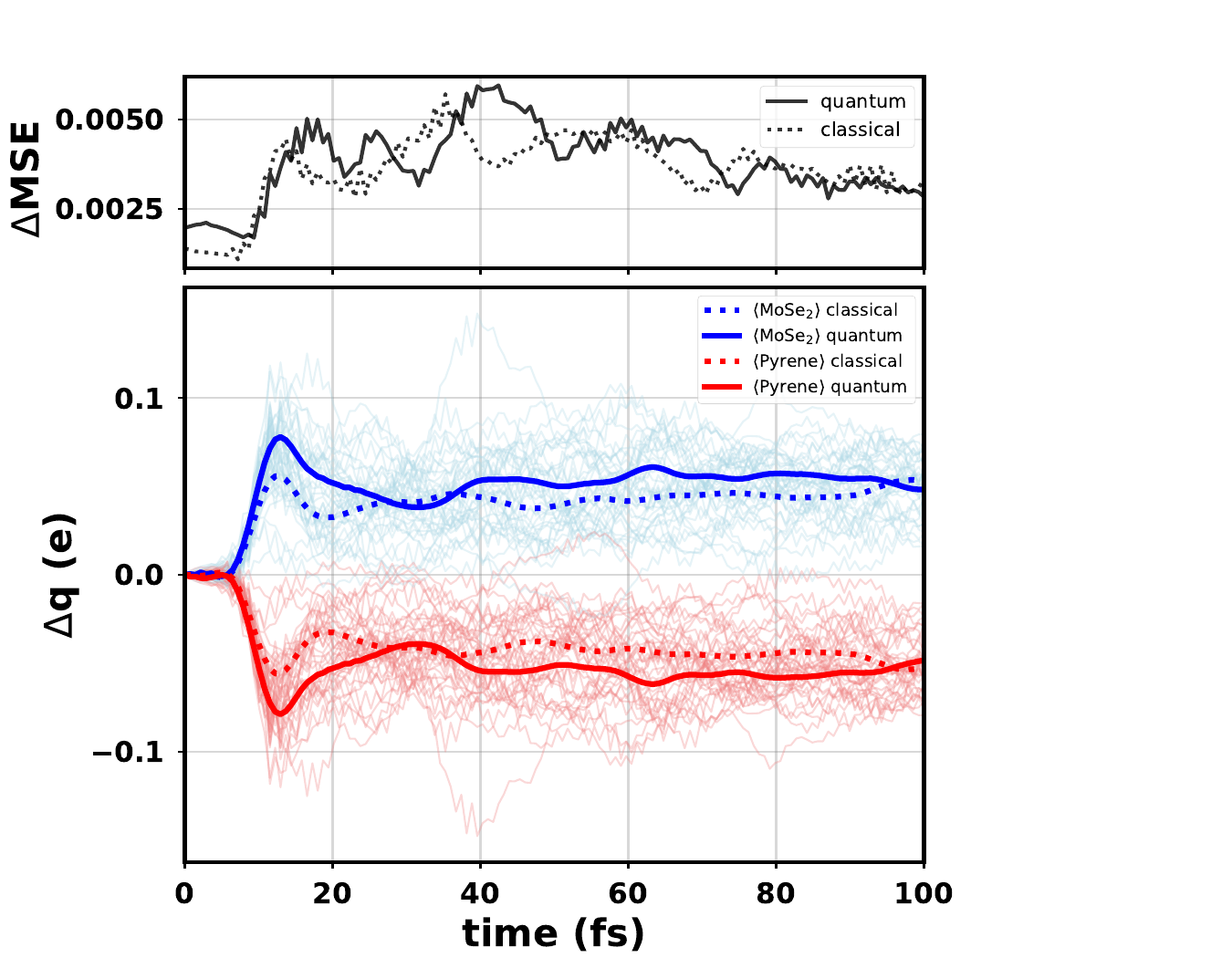}
    \caption{Time-dependent variation with respect to the ground-state value of the Bader partial charges on pyrene and MoSe$_2$ averaged over 20 snapshots in the classical (dotted lines) and quantum (solid lines) distribution smoothed using the Savitzky-Golay filter. Positive (negative) values indicate electron accumulation (depletion). The colored lines in the background represent the single trajectories for both distributions before averaging. The upper panel displays the mean-square error for both the classical (dotted line) and quantum distribution (solid line). 
    }
    \label{fig:charge}
\end{figure}

In the last part of our analysis, we focus on the partial charge distribution between the components of the interface simulated with classical and quantum nuclear distributions.
The trend observed in Fig.~\ref{fig:0K}b manifests itself also in the results of the thermalized ensembles (see Fig.~\ref{fig:charge}): In the heterostructure, the TMDC withdraws charge from the molecule excited by the pulse in resonance with its HOMO$\rightarrow$LUMO transition.
However, some differences can be noticed.
In both CD and QD, the amplitude of $\Delta q$ after the switching off of the laser (15~fs) is reduced compared to the single-trajectory case. Additionally, we do not observe a decay of the amount of charge transferred within this time window in contrast to Fig.~\ref{fig:0K}b.
In this case, there is only one noticeable difference in the results obtained with CD and QD: With the former, the maximum of $\Delta q$ at 15~fs is comparable to the $\Delta q$ observed at later times, while with the latter, it is clearly larger.

The qualitative similarity between the RT-TDDFT+Ehrenfest results for $\Delta q$ presented in this paper and those obtained in the absence of nuclear effects~\cite{jacobs+ACS2022} suggest the dominance of electronic mechanisms in the laser-induced charge transfer occurring in this system in the explored 100~fs window. 
It can be expected that phonon modes, especially related to the inorganic substrate, can come into play but only on much longer timescales (about 1~ps) compared to the one considered here.
Also, it is important to highlight the possibility of suppression of the relative impact of nuclear effects due to the intensity of the laser pulse and the resulting large energy transfer to specific phonon modes~\cite{carusojpcl2021}. Nevertheless, these findings support that the behavior of ultrafast charge transfer in this system is qualitatively robust to nuclear effects.

%%%%%%%%%%%%%%%%%%
\section{Summary and Conclusions}
To summarize, we have investigated the nuclear effects in the ultrafast dynamics of the low-dimensional hybrid interface formed by pyrene physisorbed on a MoSe$_2$ monolayer. By means of RT-TDDFT and multi-trajectory Ehrenfest dynamics, we explored the effects of the initial nuclear distribution at the time of laser excitation and those induced by the nuclear motion in a 100~fs window following this excitation. The initial nuclear distributions were obtained from AIMD with a classical treatment of the nuclei and employing a quantum thermostat that mimics nuclear quantum fluctuations, both at a temperature of 100~K.

In the analysis, we focused on the linear absorption spectra, the laser-induced electronic density current, and the charge transfer dynamics processed in terms of Bader partial charges. In the optical spectra, we found that the excitations below 3~eV, which all stem from interband transitions in MoSe$_{2}$, are less sensitive to structural changes of the system, in contrast to the sharper peak at 3.1~eV, corresponding to the HOMO$\rightarrow$LUMO transition in pyrene. 
This resonance shows a blue shift at finite temperatures, indicating that the excitation in pyrene is more sensitive to conformational changes and temperature effects, in agreement with the large amplitude motions involving the molecule in the initial nuclear distributions. The electronic current density is affected by the classical and quantum distributions as testified by the appearance of different beating patterns and decay times of the signal. The decay is fastest when considering the quantum initial nuclear distribution. This finding suggests that the decay mechanism and lifetime of the excited states could change when considering classical or quantum nuclear degrees of freedom. However, to be more conclusive in this regard, an extensive analysis is still needed.

The dynamical Bader charge analysis indicates that when excited by a pulse in resonance with its HOMO$\rightarrow$LUMO transition, pyrene acts as a donor and MoSe$_2$ as an acceptor when including all nuclear effects. This result is in agreement with the findings previously obtained without including nuclear dynamics~\cite{jacobs+ACS2022}, showing that the charge transfer is strongly dominated by electronic effects and robust to nuclear fluctuations and dynamics at short time scales. In fact, multi-trajectory Ehrenfest dynamics show that charge transfer is stabilized in comparison to a single-trajectory protocol. Of course, it should be considered that the simulations analyzed in this work cover a time window of 100~fs, in which only the vibrational modes of pyrene are activated, as the phonons of MoSe$_2$ are expected to participate on much longer timescales, on the order of a few picoseconds. The choice of a quantum or classical treatment of the nuclei affects only quantitative details of the charge distribution but not any of its qualitative features.

The findings presented in this work provide new insight into the role of nuclear effects in the ultrafast dynamics of low-dimensional hybrid interfaces excited by a resonant, femtosecond pulse. We can conclude that in the explored 100~fs time window, the dominant mechanisms in the response of the systems to the external perturbation are essentially electronic. 
In future studies, it will be interesting to employ techniques that can reliably extend the propagation time to explore the regime in which the phonons of MoSe$_2$ participate in the dynamics, too. In addition, it will be important to consider also lower laser intensities to understand whether nuclear effects may become dominant in such a case. Yet, the knowledge gained from this study offers already important information regarding the fundamental behavior of weakly-bound low-dimensional hybrid interfaces coherently interacting with radiation, showing that even when dealing with large amplitude nuclear motions and laser intensities that are not exceedingly high, the short-time ultrafast dynamics can still be dominated by electronic effects, offering guidance to the design and optimization of these systems as active components in optoelectronic applications.

%%%%%%%%%%%%%%%%%%%%%%%%%%%%%%%%%%
\section*{Acknowledgement}
This work was funded by the German Research Foundation, project number 182087777 (CRC 951), by the German Federal Ministry of Education and Research (Professorinnenprogramm III), and by the State of Lower Saxony (Professorinnen f\"ur Niedersachsen, SMART, and DyNano). The computational resources were provided by North-German Supercomputing Alliance (HLRN) through the project bep00104. M.R. acknowledges informative and fruitful discussions with Aaron Kelly about multi-trajectory Ehrenfest dynamics and computer time at MPCDF. 

%%%%%%%%%%%%%%%%%%%%%%%%%%%%%%%%%%%%
\section*{Data Availability Statement}
The data that support the findings of this study are available from the corresponding authors upon reasonable request.
%%%%%%%%%%%%%%%%%%%%%%%%%%%%%%%%%%%%
%\newpage
\section*{References}
\bibliographystyle{iopart-num}
%\bibliography{bib}

\providecommand{\newblock}{}

\end{document}